 \documentclass[final,5p,times,twocolumn]{elsarticle}

\usepackage{amssymb}
\usepackage{amsthm}

\usepackage{lineno,hyperref}
\usepackage{cleveref}
\usepackage{mathptmx}
\journal{Nuclear Instruments and Methods in Physics Research A}

\begin{document}

\begin{frontmatter}



\title{RF injector design studies for the trailing witness bunch for a plasma-based user facility}

\author[label1]{ A. Giribono }\fnref{myfootnote}
\author[label3]{ A. Bacci}
\author[label1]{ E. Chiadroni }
\author[label2]{ A. Cianchi }
\author[label1]{ M. Croia }
\author[label1]{ M. Ferrario }
\author[label1]{ A. Marocchino }
\author[label3]{ V. Petrillo }
\author[label1]{ R. Pompili }
\author[label1]{ S. Romeo}
\author[label3]{ M. Rossetti Conti}
\author[label3]{ A.R. Rossi}
\author[label1]{ C. Vaccarezza}

\address[label1]{INFN-LNF, Via Enrico Fermi 40, 00044 Frascati Rome, Italy}
\address[label2]{INFN-Tor Vergata University, Via Ricerca Scientifica 1, 00133 Rome, Italy}
\address[label3]{INFN-MI, Via Celoria 16, 20133 Milan, Italy}

\fntext[myfootnote]{Email address: anna.giribono@lnf.infn.it}




\begin{abstract}

The interest in plasma-based accelerators as drivers of user facilities is growing worldwide thanks to its compactness and reduced costs. In this context the EuPRAXIA collaboration is preparing a conceptual design report for a multi-GeV plasma-based accelerator with outstanding electron beam quality to pilot, among several applications, the operation of an X-ray FEL, the most demanding in terms of beam brightness. Intense beam dynamics studies have been performed to provide a reliable working point for the RF injector to generate a high-brightness trailing witness bunch suitable in external injection schemes, both in particle beam and laser driven plasma wakefield acceleration. A case of interest is the generation of a witness beam with 1 GeV energy, less than 1 mm-mrad slice emittance and 30 pC in 10 fs FWHM bunch length, which turns into 3 kA peak current at the undulator entrance. The witness beam has been successfully compressed down to 10 fs in a conventional SPARC-like photo-injector
and boosted up to 500 MeV in an advanced high-gradient X-band linac reaching the plasma entrance with 3 kA peak current and the following RMS values: 0.06$\%$ energy spread, 0.5 mm-mrad transverse normalised emittance and a focal spot down to 1 $\mu m$. RF injector studies are here presented with the aim to satisfy the EuPRAXIA requests for the Design Study of a plasma-based user facility.

\end{abstract}


\end{frontmatter}


\section{Introduction}

The EuPRAXIA collaboration \cite{item300} is preparing a conceptual design report for a multi-GeV plasma-based accelerator with outstanding beam quality  to drive a user facility for several applications (photon science, HEP, radiation sources, ... ). Many configurations for the plasma accelerating stage are under study including the particle beam or laser driven plasma wakefield acceleration (PWFA and LWFA, respectively) with external injection from an RF-generated electron beam. In this case the RF injector will provide at the plasma entrance an electron beam whose parameters are the ones requested at the undulator entrance, except for the energy. Indeed, for the successful operation of a plasma-based user facility, the plasma structure should not introduce any degradation of the beam quality but only boost the energy. 

Driven by the most demanding user application, i.e. the operation of an X-ray Free-Electron Laser, which requires ultra-high brightness electron beams (peak current I$_{peak}$ $\simeq$ 3 kA, rms transverse normalised emittance $\epsilon_{n_{x,y}}$ $<$ 1 mm-mrad, 1 GeV energy and rms energy spread $\Delta\gamma/\gamma$ $<$ 0.1$\%$), a preliminary parameter list has been conceived for different plasma acceleration schemes and working points. In particular, an electron beam being injected in the undulator with 1 GeV energy, less than 1 mm-mrad slice emittance (slice length $l_{slice}$ = 0.75 $\mu m$), 30 pC charge in 10 fs FWHM bunch length, which turns into 3 kA peak current (total charge divided by FWHM bunch length), is required to enable the process of self-amplified spontaneous emission FEL at $\lambda_r$ $\simeq$ 3 nm with K = 1.

A proposal of an optimised RF injector is documented in the following sections with the aim to satisfy the EuPRAXIA requests for the Design Study of a plasma-based user facility independently by the driving mechanism.

\begin{figure*}[!ht]
	\begin{center}
		\includegraphics[width=0.9\textwidth]{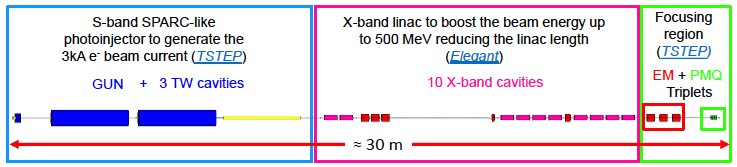}    
		\caption{Layout of the SPARC-like high brightness S-band photo-injector (blue box) consisting of a 1.6 cell UCLA/BNL type SW RF gun, equipped with a copper photo cathode and an emittance compensation solenoid, followed by three TW SLAC type sections; other two compensation solenoids surround the first and the second S-band cavities for the operation in the velocity bunching scheme. The pink box wraps the X-band linac, including matching and diagnostics section (an electro-magnet (EM) quadrupole triplet, followed by a 4 m drift). The green box includes the matching and final focus transfer line to the plasma interface, composed by an electro-magnet (EM) quadrupole triplet and a permanent magnet quadrupole (PMQ) triplet.}
		\label{fig:lin_la}
	\end{center}
\end{figure*}

\section{RF Injector Scheme Layout}

A possible layout for the conventional RF injector is about 28 m long, including matching and diagnostics sections, and is composed as follows, see \figurename~\ref{fig:lin_la}:
\begin{itemize}
	\item an S-band photo-injector (a 1.6 cell standing wave RF gun and three TW sections) to generate a 3 kA beam current  through RF compression in the first two TW structures, operating in the velocity bunching regime, 
	\item an X-band linac to boost the beam energy up to about 500 MeV,
	\item a final strong focusing system to match the beam transversally at the plasma entrance.
\end{itemize}

\begin{figure}
	\begin{center}
		\includegraphics[width=1\columnwidth]{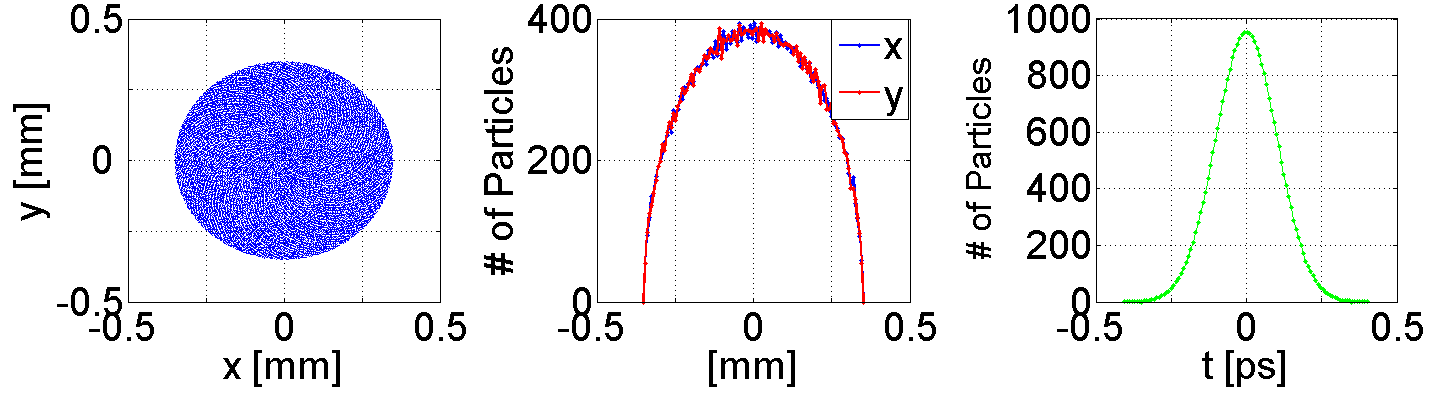}    
		\caption{Charge distribution at cathode surface produced by the photo-cathode laser pulse as obtained with 2D Tstep simulations. }
		\label{fig:cathode}
	\end{center}
\end{figure}

The performance of such injector is reported in the following sections. 

\subsection{The Photo-Injector}
The SPARC-like photo-injector \cite{item29},operating at 2.856 GHz, is composed of a 1.6 cell UCLA/BNL S-band RF gun type equipped with a copper photo cathode and an emittance compensation solenoid, followed by three TW SLAC type S-band sections (3-meter long each) \cite{item14}; other two compensation solenoids surround the first two S-band cavities for the operation in the velocity bunching scheme. The photo-injector has been optimised to provide at the X-band entrance a 30 pC electron beam with transverse normalised emittance, $\epsilon_{n_{x,y}}$, lower than 1 mm-mrad, FWHM bunch length, $\tau_z$ $\simeq$ 3 $\mu m$ and energy of $\simeq$ 100 MeV.

The beam line matching foresees a proper set of the emittance compensation solenoids and of the S-band cavity gradients in the so-called  velocity bunching scheme, according to the invariant envelope criteria \cite{item81,item39}. The first and second TW sections operate far from the crest in the velocity bunching regime to fully compress the witness beam length down to 3 $\mu m$ (10 fs) FWHM, which corresponds to a peak current of 3 kA, and the third section operates almost on crest maximize the energy gain and freeze its phase space quality. 

\subsection{The X-band Linac}
The X-band RF linac, operating at 11.424 GHz, is composed of ten accelerating sections and can boost the electron beam up to $\approx$ 550 MeV energy. Here the accelerating gradient can be set up to 90 MV/m on average, allowing enough margin for the off crest operation that minimise the final beam energy spread. The X-band accelerating structures are 0.5 m long and consist of $\simeq$ 8.3 mm long cells. The beam line matching foresees to properly set magnetic elements and accelerating cavities, including wake fields in the X-band sections.

\section{Beam dynamics simulations}
Intense beam dynamics studies have been performed to provide a reliable working point for the injector to drive a witness bunch suitable for external injection schemes, both in PWFA and LWFA.  The beam parameters at the plasma entrance have been optimised for the 30 pC nominal electron beam with 30 k macro particles in the single bunch operation with the following characteristics: 3 kA peak current, $I_{peak}$,  energy spread $\Delta\gamma/\gamma$ $<$ 0.1$\%$, transverse normalised emittance $\epsilon_{n_{x,y}}$ $<$ 1 mm-mrad and a final energy of about 500 MeV for both schemes.

\subsection{The Photo-injector}
The beam dynamics in the photo-injector has been simulated with the multi-particles code Tstep \cite{item6a}, which takes into account the space charge effects and the thermal emittance.
%

The  bunch distribution at the cathode surface strongly impacts on the optimisation process of final parameters at the injector exit. Here the choice has been to explore the blow-out regime by adopting a photo-cathode laser with a gaussian longitudinal profile of RMS duration $\sigma_z$ $\simeq$ 120 $mu m$ and a transverse uniform distribution of RMS spot size $\sigma_r$ = 175 $\mu m$. The \figurename~\ref{fig:cathode} shows the shaped charge distribution at the cathode surface produced by such laser pulse as obtained with 2D Tstep simulations; the cylindrical symmetry of the beam allows us to adopt a 2D model which requires a reasonable number of particles and mesh points, and so computational time.
\begin{figure}
	\begin{center}
		\includegraphics[width=1\columnwidth]{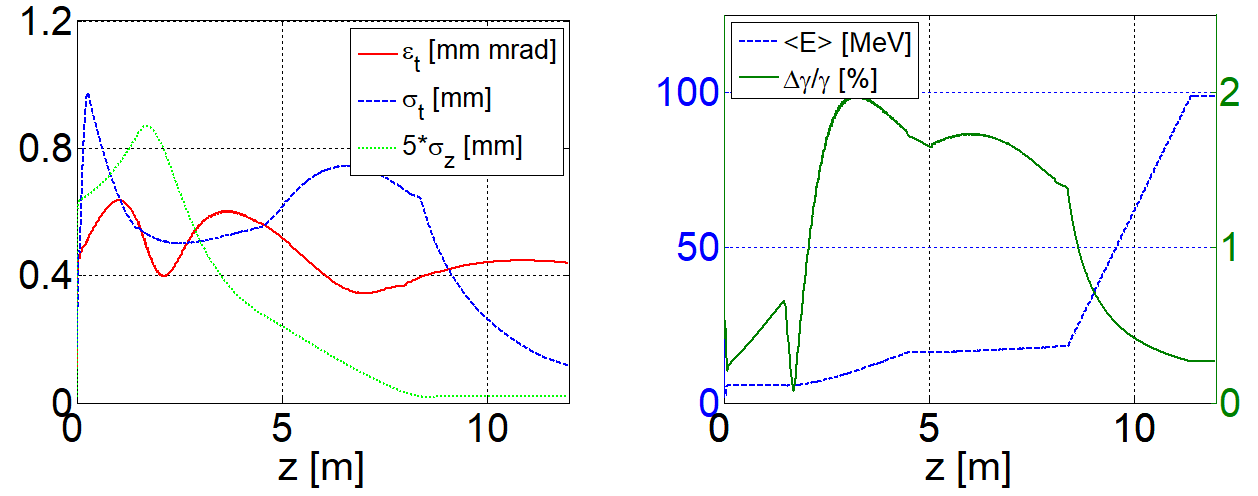}    
		\caption{Simulation results for the photo-injector: evolution along the injector of the electron beam transverse normalised emittance ($\epsilon_{n_{x,y}}$ red line), spot size ($\sigma_{x,y}$ blue dash-dot line) and longitudinal bunch length ($\sigma_z$ green dashed line) as obtained with the Tstep code.}
		\label{fig:inj_evo}
	\end{center}
\end{figure}

\begin{figure}
	\begin{center}
		\includegraphics[width=1\columnwidth]{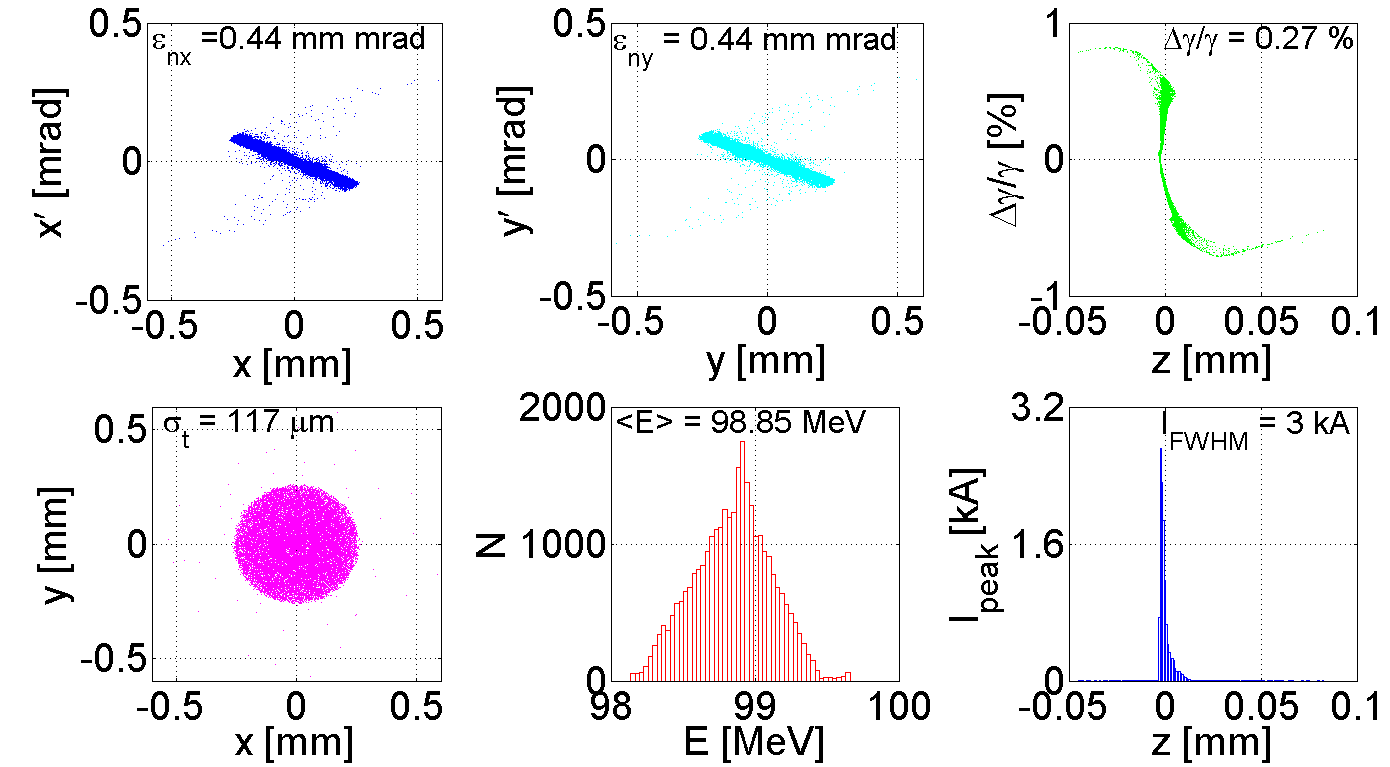}    
		\caption{Upper plots: transverse (x and y) and longitudinal phase spaces; lower plots: transverse distribution and energy and current profile. The results are output from TStep code at the photo-injector.}
		\label{fig:inj_phsp}
	\end{center}
\end{figure}

The peak RF gun accelerating field is $E_{acc}\simeq$ 120 MV/m and a slightly dephasing between the field and the beam allows to maximise the energy gain in this part. Then, the first two S-band structures operate at $E_{acc}$ = 20.0 MV/m and the third one at $E_{acc}$ = 28.0 MV/m on average.\\ The velocity bunching scheme in the first two S-band cavities is adopted to shorten the beam length from 350 to $\simeq$ 5 $\mu m$, i.e. dephasing of  few degrees with respect to the zero crossing. Finally, the emittance minimisation is obtained setting the gun solenoid at $\simeq$ 3 kG, the one surrounding the first and second S-band cavities at respectively $\simeq$ 0.32 kG and 0.50 kG. A slightly off-crest operation of the third S-band cavity further reduces the energy spread at the injector exit. 
In this configuration the design electron beam parameters at the photo-injector exit, listed in \tablename~\ref{tab:table_inj}, are: E = 98.85 MeV, $\epsilon_{n_{x,y}}$ = 0.44 mm mrad, $\sigma_{x,y}$ = 117 $\mu m$, $\Delta\gamma/\gamma$ = 0.27 \%, $\sigma_z$ = 5.6 $\mu m$. The longitudinal profile of the beam exiting the photo-injector is typical of the velocity bunching process showing a spike of current on the head of the bunch and a long tail and a  $\tau_{z}$ $\simeq$ 3 $\mu m$. The simulation results are shown in  \figurename~\ref{fig:inj_evo} and \figurename~\ref{fig:inj_phsp} for a 30 pC electron beam with final energy of 98.8 MeV. The \figurename~\ref{fig:inj_evo} (left plot) shows the evolution of the transverse normalised emittance (red line), spot size (blue line) and longitudinal bunch length (green line), while the right plot illustrates the energy (blue line) and energy spread (green line) from the cathode down to the photo-injector exit as obtained with the Tstep code. Further longitudinal and transverse phase spaces are reported at the photo-injector exit as obtained with Tstep in \figurename~\ref{fig:inj_phsp}. 
\begin{table}
	\centering
	\caption{ Simulated parameters of the 30 pC electron beam at photo-injector exit and plasma entrance as result from beam dynamics studies.}
	\vskip 0.1 in
	\begin{tabular}{|l|c|c|} \hline
		&  @Photo-injector & @Plasma  \\
		&  Exit & Entrance \\
		\hline
		\hline
		E [MeV]   & 98.85  & 518.00\\
		$\frac{\Delta\gamma}{\gamma}$ [\%] & 0.27 & 0.06\\
		$\sigma_{x,y}$  $[\mu m]$ &  117.00  & 1.00\\
		$\epsilon_{n_{x,y}}$  [mm mrad] & 0.44  & 0.45 - 0.47\\
		$\beta_{x,y}$  [m] & 6.1 & 0.002\\
		$\alpha_{x,y}$ & 2.1 & 0.00\\
		$\sigma_{z}$  $[\mu m]$  & 5.63   & 6.00 \\
		$\tau_{z}$  $[\mu m]$ &  3.00 & 3.00\\
		$I_{peak} $ [kA] &3.00   &  3.00\\
		\hline
	\end{tabular}
	\label{tab:table_inj}
\end{table}

\subsection{The X-band Linac}
Electron beam dynamics simulations in the X-band booster linac have been performed with the Elegant code \cite{item6b} that includes the wakefields generated by the electron beam inside the accelerating structures together with the longitudinal space charge and the coherent and incoherent synchrotron radiation effects in the bending magnets. The code computes wake fields making the convolution of the specified time-dependent moment of the particle distribution with the Green wake function, defined as the time response of a system to a unit impulse. The longitudinal wake function is expressed in \emph{Volt/Coulomb}, while the transverse wake function as \emph{Volt/(Coulomb meter)} in order to be independent on the offset of the driving charge.

The asymptotic values of the longitudinal and transverse wake fields have been calculated for the X-band structures according to \cite{item20}
\begin{equation}
W_{0||}(s) \approx \frac{Z_0 c}{\pi a^2}exp(-\sqrt{\frac{s}{s_1}})\quad \frac{V}{C m} 
\label{W_long}
\end{equation}
with s$_1$ = 0.41 $\frac{a^{1.8}g^{1.6}}{L^{2.4}}$

 \begin{figure}
 	\begin{center}
 		\includegraphics[width=1\columnwidth]{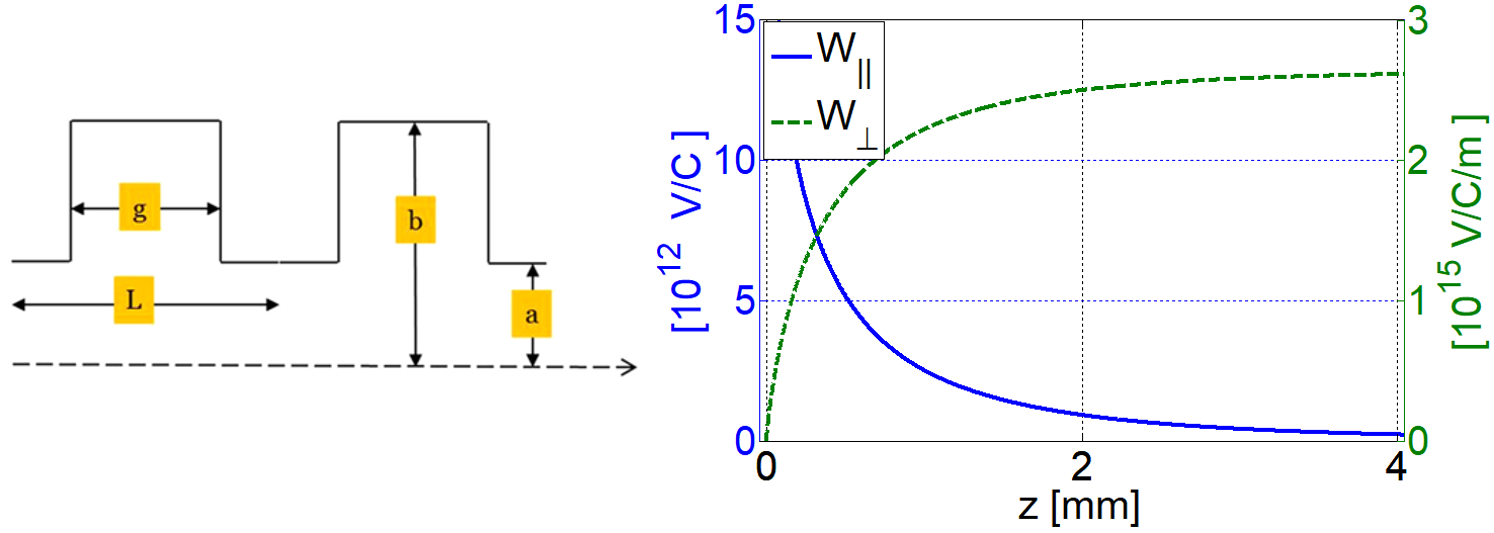}    
 		\caption{Left: Pill box cavity model considered for the wake fields calculations. Right: Longitudinal and transverse short-range wake function curves integrated over one cell for the X-band accelerating structure.}
 		\label{fig:wake_2}
 	\end{center}
 \end{figure}

\begin{equation}
W_{0\perp}(s) \approx \frac{4Z_0cs_2}{\pi a^4}[1-(1+\sqrt{\frac{s}{s_2}})exp(-\sqrt{\frac{s}{s_2}})]\quad  \frac{V}{C m^2} 
\label{W_tras}
\end{equation}
with s$_2$ = 0.17 $\frac{a^{1.79}g^{0.38}}{L^{1.17}}$

\vspace{5mm}
where Z$_0$ is the free space impedance, c is the light velocity and a, L, b and g are defined as in \figurename~\ref{fig:wake_2}. The calculated Green function wakes used in Elegant for the X-band accelerating structures are reported in \figurename~\ref{fig:wake_2}.

\begin{figure}
	\begin{center}
		\includegraphics[width=0.9\columnwidth]{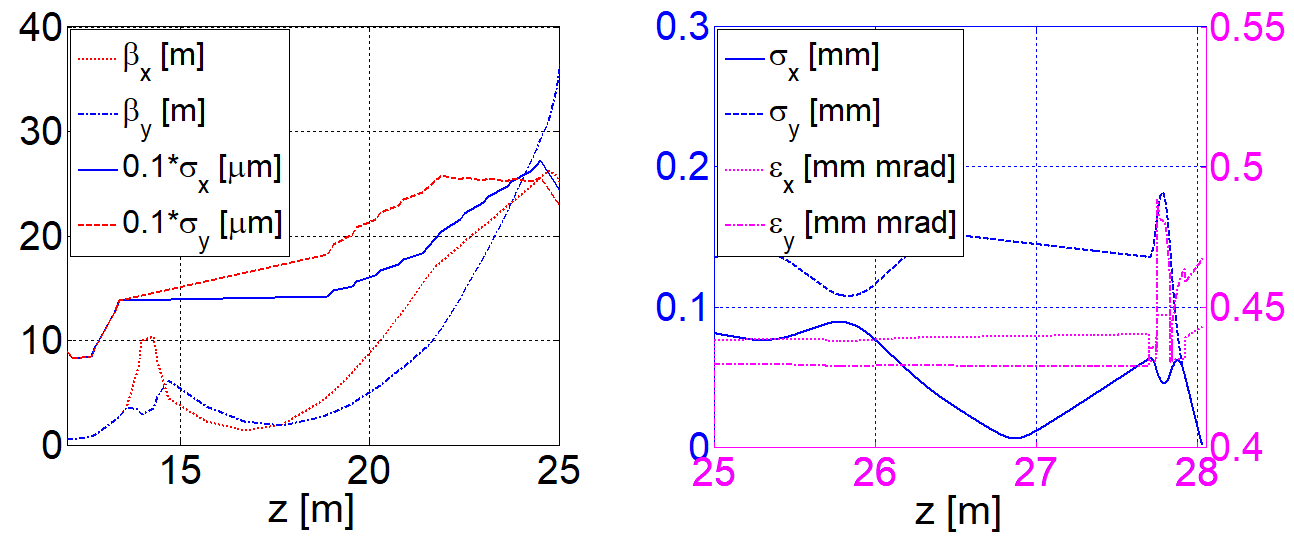}    
		\caption{Simulation results for the X-band linac and matching section: evolution of the simulated Twiss parameters from the photo-injector exit down to the final focusing region entrance as obtained with Elegant (left side). The beam dynamics in the final focusing region is highlighted on the right, showing the transverse spot size and normalised emittance as obtained with TStep.}
		\label{fig:lin_evo}
	\end{center}
\end{figure}

The working point has been optimised to provide at the plasma entrance an electron beam energy with the simulated parameters listed in \tablename~\ref{tab:table_inj}. Accelerating gradients set at 80 MV/m average gradient, together with the proper off crest operation, i.e. dephasing of $\pm$15 degrees with respect to the maximum RF accelerating field, allow to reach energy spread values of $\simeq$ 0.06 \%. The evolution of simulated Twiss parameters from the photo-injector exit down to the final focusing region entrance as obtained with Elegant in shown in \figurename~\ref{fig:lin_evo} (left side).
It has to be mentioned that for the adopted working point the transition energy between the space charge and emittance dominated regimes is higher than 100 MeV; nevertheless only a slightly difference can be observed extending the multi particle simulation to the entrance of the downstream strong focusing region.

\subsection{The strong focusing system}

The strong focusing system, consisting in an electro-magnet and a permanent quadrupole triplet, has been designed to match the beam transversely at the plasma entrance with a beta function of about 1 mm, corresponding to a transverse spot size $\sigma_{x,y}$ $\simeq$ 1 $\mu m$ at the plasma interface, as requested by the transverse matching condition, being the plasma wavelength $\lambda_p$ $\simeq$ 334 $\mu m$ for the target plasma background density $n_p$ = $10^{16}$ $cm^{-3}$.

The beam dynamics in this part of the machine has been simulated with Tstep to evaluate the transverse emittance and longitudinal distribution dilution due to the space charge contribution in the strong focusing region. The transverse spot size and normalised emittance in the focusing region as obtained with TStep are reported in \figurename~\ref{fig:lin_evo} (right side).

\begin{figure}
	\begin{center}
		\includegraphics[width=1\columnwidth]{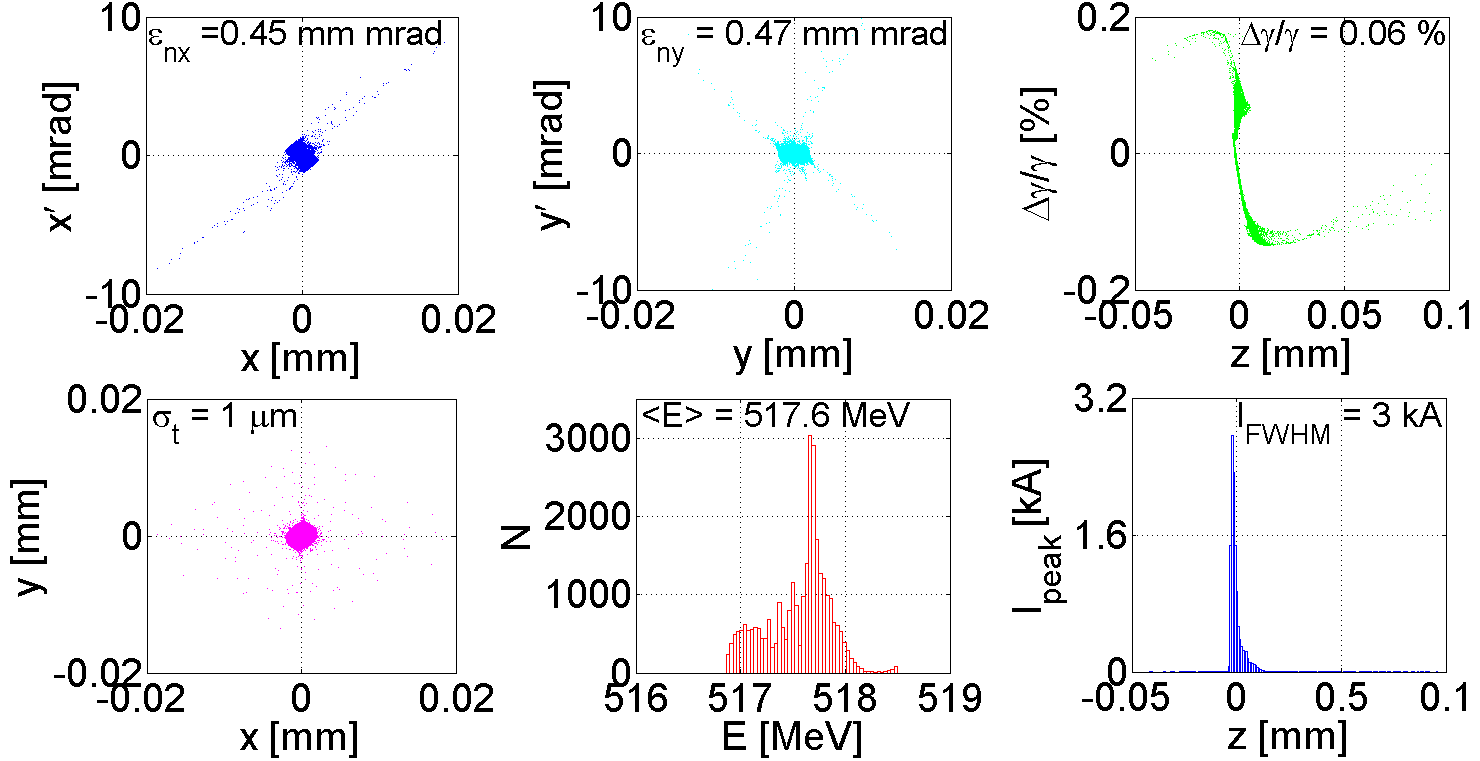}    
		\caption{Upper plots: transverse (x and y) and longitudinal phase spaces; lower plots: transverse distribution and energy and current profile. The results are output from TStep code at the plasma entrance.}
		\label{fig:foc_phsp}
	\end{center}
\end{figure}

The simulated longitudinal and transverse phase spaces are reported at the plasma entrance as obtained with Tstep in \figurename~\ref{fig:foc_phsp}. Both transverse and longitudinal phase spaces have been preserved as shown in \figurename~\ref{fig:foc_phsp}: the final energy spread is 0.06 $\%$ and the final FWHM bunch length is $\simeq$  3.0 $\mu m$ (10 fs), which corresponds to a peak current of 3 kA, with 0.46 mm-mrad transverse normalised emittance and $\sigma_{x,y}$ = 1 $\mu m$.

\section{Conclusions}
Design studies for an RF injector as driver for a plasma-based accelerator have been presented with promising results in terms of electron beam phase space quality, making it an effective candidate to pilot the EuPRAXIA plasma-based facility. In particular, the proposed RF injector scheme enables the generation of a 3 kA trailing bunch, a milestone of the EuPRAXIA European Design Study, with following parameters at the plasma accelerator entrance: 518 MeV energy, $\tau_z$ is $\simeq$ 3.0 $\mu m$ (10 fs), $I_{peak}$ $\simeq$ 3.0 kA, $\Delta\gamma/\gamma$  $\simeq$ 0.06 $\%$, $\epsilon_{n_{x,y}}$ = 0.46 mm-mrad and $\sigma_{x,y}$ = 1 $\mu m$. The beam dynamics of the trailing bunch, from the cathode up to the FEL, is now under study by means of start to end simulations, considering both laser and particle driven plasma wakefield acceleration. In more details, the generation of both driver and trailing beams directly on the cathode, through a proper phase space manipulation in the rest of the RF injector, is being studied in case of particle driven scheme for the plasma-stage. Further, injector sensitivity analysis are ongoing to evaluate the robustness and reliability of this working point for efficient acceleration in the plasma and finally to drive the FEL emission.

\section*{Acknowledgment}
This work was supported by the European Union's Horizon 2020 research and innovation programme under grant agreement No. 653782.






\end{document}